\documentclass[twocolumn,letterpaper,aps,prl,longbibliography,superscriptaddress,showpacs,floatfix]{revtex4-1}

\usepackage{graphicx}	
\usepackage{amsmath}
\usepackage{xspace}

\newcommand{\pt}{\mbox{$p_T$}\xspace}
\newcommand{\Npart}{\mbox{$N_{\rm part}$}\xspace}

\newcommand{\Nch}{\mbox{$N_{\rm ch}$}\xspace}

\newcommand{\sqsn}{\mbox{$\sqrt{s_{_{NN}}}$}\xspace}
\newcommand{\auau}{\mbox{Au$+$Au}\xspace}
\begin{document}


\title{Measurement of higher cumulants of net-charge multiplicity 
distributions in Au$+$Au collisions at $\sqrt{s_{_{NN}}}=7.7$--200 GeV}

\newcommand{\abilene}{Abilene Christian University, Abilene, Texas 79699, USA}
\newcommand{\augie}{Department of Physics, Augustana University, Sioux Falls, South Dakota 57197, USA}
\newcommand{\banaras}{Department of Physics, Banaras Hindu University, Varanasi 221005, India}
\newcommand{\barc}{Bhabha Atomic Research Centre, Bombay 400 085, India}
\newcommand{\baruch}{Baruch College, City University of New York, New York, New York, 10010 USA}
\newcommand{\bnlcoll}{Collider-Accelerator Department, Brookhaven National Laboratory, Upton, New York 11973-5000, USA}
\newcommand{\bnlphys}{Physics Department, Brookhaven National Laboratory, Upton, New York 11973-5000, USA}
\newcommand{\caucr}{University of California-Riverside, Riverside, California 92521, USA}
\newcommand{\charlesczech}{Charles University, Ovocn\'{y} trh 5, Praha 1, 116 36, Prague, Czech Republic}
\newcommand{\chonbuk}{Chonbuk National University, Jeonju, 561-756, Korea}
\newcommand{\ciae}{Science and Technology on Nuclear Data Laboratory, China Institute of Atomic Energy, Beijing 102413, People's Republic of China}
\newcommand{\cns}{Center for Nuclear Study, Graduate School of Science, University of Tokyo, 7-3-1 Hongo, Bunkyo, Tokyo 113-0033, Japan}
\newcommand{\colorado}{University of Colorado, Boulder, Colorado 80309, USA}
\newcommand{\columbia}{Columbia University, New York, New York 10027 and Nevis Laboratories, Irvington, New York 10533, USA}
\newcommand{\czechtech}{Czech Technical University, Zikova 4, 166 36 Prague 6, Czech Republic}
\newcommand{\dapnia}{Dapnia, CEA Saclay, F-91191, Gif-sur-Yvette, France}
\newcommand{\debrecen}{Debrecen University, H-4010 Debrecen, Egyetem t{\'e}r 1, Hungary}
\newcommand{\elte}{ELTE, E{\"o}tv{\"o}s Lor{\'a}nd University, H-1117 Budapest, P{\'a}zm{\'a}ny P.~s.~1/A, Hungary}
\newcommand{\ewha}{Ewha Womans University, Seoul 120-750, Korea}
\newcommand{\fit}{Florida Institute of Technology, Melbourne, Florida 32901, USA}
\newcommand{\fsu}{Florida State University, Tallahassee, Florida 32306, USA}
\newcommand{\gsu}{Georgia State University, Atlanta, Georgia 30303, USA}
\newcommand{\hanyang}{Hanyang University, Seoul 133-792, Korea}
\newcommand{\hiroshima}{Hiroshima University, Kagamiyama, Higashi-Hiroshima 739-8526, Japan}
\newcommand{\ihepprot}{IHEP Protvino, State Research Center of Russian Federation, Institute for High Energy Physics, Protvino, 142281, Russia}
\newcommand{\illuiuc}{University of Illinois at Urbana-Champaign, Urbana, Illinois 61801, USA}
\newcommand{\inrras}{Institute for Nuclear Research of the Russian Academy of Sciences, prospekt 60-letiya Oktyabrya 7a, Moscow 117312, Russia}
\newcommand{\instpasczech}{Institute of Physics, Academy of Sciences of the Czech Republic, Na Slovance 2, 182 21 Prague 8, Czech Republic}
\newcommand{\isu}{Iowa State University, Ames, Iowa 50011, USA}
\newcommand{\jaea}{Advanced Science Research Center, Japan Atomic Energy Agency, 2-4 Shirakata Shirane, Tokai-mura, Naka-gun, Ibaraki-ken 319-1195, Japan}
\newcommand{\jinrdubna}{Joint Institute for Nuclear Research, 141980 Dubna, Moscow Region, Russia}
\newcommand{\jyvaskyla}{Helsinki Institute of Physics and University of Jyv{\"a}skyl{\"a}, P.O.Box 35, FI-40014 Jyv{\"a}skyl{\"a}, Finland}
\newcommand{\kek}{KEK, High Energy Accelerator Research Organization, Tsukuba, Ibaraki 305-0801, Japan}
\newcommand{\korea}{Korea University, Seoul, 136-701, Korea}
\newcommand{\kurchatov}{National Research Center ``Kurchatov Institute", Moscow, 123098 Russia}
\newcommand{\kyoto}{Kyoto University, Kyoto 606-8502, Japan}
\newcommand{\labllr}{Laboratoire Leprince-Ringuet, Ecole Polytechnique, CNRS-IN2P3, Route de Saclay, F-91128, Palaiseau, France}
\newcommand{\lahorelums}{Physics Department, Lahore University of Management Sciences, Lahore 54792, Pakistan}
\newcommand{\lawllnl}{Lawrence Livermore National Laboratory, Livermore, California 94550, USA}
\newcommand{\losalamos}{Los Alamos National Laboratory, Los Alamos, New Mexico 87545, USA}
\newcommand{\lpc}{LPC, Universit{\'e} Blaise Pascal, CNRS-IN2P3, Clermont-Fd, 63177 Aubiere Cedex, France}
\newcommand{\lund}{Department of Physics, Lund University, Box 118, SE-221 00 Lund, Sweden}
\newcommand{\maryland}{University of Maryland, College Park, Maryland 20742, USA}
\newcommand{\mass}{Department of Physics, University of Massachusetts, Amherst, Massachusetts 01003-9337, USA}
\newcommand{\michigan}{Department of Physics, University of Michigan, Ann Arbor, Michigan 48109-1040, USA}
\newcommand{\muenster}{Institut f\"ur Kernphysik, University of Muenster, D-48149 Muenster, Germany}
\newcommand{\muhlenberg}{Muhlenberg College, Allentown, Pennsylvania 18104-5586, USA}
\newcommand{\myongji}{Myongji University, Yongin, Kyonggido 449-728, Korea}
\newcommand{\nagasaki}{Nagasaki Institute of Applied Science, Nagasaki-shi, Nagasaki 851-0193, Japan}
\newcommand{\nara}{Nara Women's University, Kita-uoya Nishi-machi Nara 630-8506, Japan}
\newcommand{\natmephi}{National Research Nuclear University, MEPhI, Moscow Engineering Physics Institute, Moscow, 115409, Russia}
\newcommand{\newmex}{University of New Mexico, Albuquerque, New Mexico 87131, USA}
\newcommand{\nmsu}{New Mexico State University, Las Cruces, New Mexico 88003, USA}
\newcommand{\ohio}{Department of Physics and Astronomy, Ohio University, Athens, Ohio 45701, USA}
\newcommand{\ornl}{Oak Ridge National Laboratory, Oak Ridge, Tennessee 37831, USA}
\newcommand{\orsay}{IPN-Orsay, Univ. Paris-Sud, CNRS/IN2P3, Universit{\'e} Paris-Saclay, BP1, F-91406, Orsay, France}
\newcommand{\peking}{Peking University, Beijing 100871, People's Republic of China}
\newcommand{\pnpi}{PNPI, Petersburg Nuclear Physics Institute, Gatchina, Leningrad region, 188300, Russia}
\newcommand{\riken}{RIKEN Nishina Center for Accelerator-Based Science, Wako, Saitama 351-0198, Japan}
\newcommand{\rikjrbrc}{RIKEN BNL Research Center, Brookhaven National Laboratory, Upton, New York 11973-5000, USA}
\newcommand{\rikkyo}{Physics Department, Rikkyo University, 3-34-1 Nishi-Ikebukuro, Toshima, Tokyo 171-8501, Japan}
\newcommand{\saispbstu}{Saint Petersburg State Polytechnic University, St.~Petersburg, 195251 Russia}
\newcommand{\saopaulo}{Universidade de S{\~a}o Paulo, Instituto de F\'{\i}sica, Caixa Postal 66318, S{\~a}o Paulo CEP05315-970, Brazil}
\newcommand{\seoulnat}{Department of Physics and Astronomy, Seoul National University, Seoul 151-742, Korea}
\newcommand{\stonybrkc}{Chemistry Department, Stony Brook University, SUNY, Stony Brook, New York 11794-3400, USA}
\newcommand{\stonycrkp}{Department of Physics and Astronomy, Stony Brook University, SUNY, Stony Brook, New York 11794-3800, USA}
\newcommand{\tenn}{University of Tennessee, Knoxville, Tennessee 37996, USA}
\newcommand{\titech}{Department of Physics, Tokyo Institute of Technology, Oh-okayama, Meguro, Tokyo 152-8551, Japan}
\newcommand{\tsukuba}{Center for Integrated Research in Fundamental Science and Engineering, University of Tsukuba, Tsukuba, Ibaraki 305, Japan}
\newcommand{\vandy}{Vanderbilt University, Nashville, Tennessee 37235, USA}
\newcommand{\waseda}{Waseda University, Advanced Research Institute for Science and Engineering, 17  Kikui-cho, Shinjuku-ku, Tokyo 162-0044, Japan}
\newcommand{\weizmann}{Weizmann Institute, Rehovot 76100, Israel}
\newcommand{\wigner}{Institute for Particle and Nuclear Physics, Wigner Research Centre for Physics, Hungarian Academy of Sciences (Wigner RCP, RMKI) H-1525 Budapest 114, POBox 49, Budapest, Hungary}
\newcommand{\yonsei}{Yonsei University, IPAP, Seoul 120-749, Korea}
\newcommand{\zagreb}{University of Zagreb, Faculty of Science, Department of Physics, Bijeni\v{c}ka 32, HR-10002 Zagreb, Croatia}
\affiliation{\abilene}
\affiliation{\augie}
\affiliation{\banaras}
\affiliation{\barc}
\affiliation{\baruch}
\affiliation{\bnlcoll}
\affiliation{\bnlphys}
\affiliation{\caucr}
\affiliation{\charlesczech}
\affiliation{\chonbuk}
\affiliation{\ciae}
\affiliation{\cns}
\affiliation{\colorado}
\affiliation{\columbia}
\affiliation{\czechtech}
\affiliation{\dapnia}
\affiliation{\debrecen}
\affiliation{\elte}
\affiliation{\ewha}
\affiliation{\fit}
\affiliation{\fsu}
\affiliation{\gsu}
\affiliation{\hanyang}
\affiliation{\hiroshima}
\affiliation{\ihepprot}
\affiliation{\illuiuc}
\affiliation{\inrras}
\affiliation{\instpasczech}
\affiliation{\isu}
\affiliation{\jaea}
\affiliation{\jinrdubna}
\affiliation{\jyvaskyla}
\affiliation{\kek}
\affiliation{\korea}
\affiliation{\kurchatov}
\affiliation{\kyoto}
\affiliation{\labllr}
\affiliation{\lahorelums}
\affiliation{\lawllnl}
\affiliation{\losalamos}
\affiliation{\lpc}
\affiliation{\lund}
\affiliation{\maryland}
\affiliation{\mass}
\affiliation{\michigan}
\affiliation{\muenster}
\affiliation{\muhlenberg}
\affiliation{\myongji}
\affiliation{\nagasaki}
\affiliation{\nara}
\affiliation{\natmephi}
\affiliation{\newmex}
\affiliation{\nmsu}
\affiliation{\ohio}
\affiliation{\ornl}
\affiliation{\orsay}
\affiliation{\peking}
\affiliation{\pnpi}
\affiliation{\riken}
\affiliation{\rikjrbrc}
\affiliation{\rikkyo}
\affiliation{\saispbstu}
\affiliation{\saopaulo}
\affiliation{\seoulnat}
\affiliation{\stonybrkc}
\affiliation{\stonycrkp}
\affiliation{\tenn}
\affiliation{\titech}
\affiliation{\tsukuba}
\affiliation{\vandy}
\affiliation{\waseda}
\affiliation{\weizmann}
\affiliation{\wigner}
\affiliation{\yonsei}
\affiliation{\zagreb}
\author{A.~Adare} \affiliation{\colorado} 
\author{S.~Afanasiev} \affiliation{\jinrdubna} 
\author{C.~Aidala} \affiliation{\losalamos} \affiliation{\mass} \affiliation{\michigan} 
\author{N.N.~Ajitanand} \affiliation{\stonybrkc} 
\author{Y.~Akiba} \affiliation{\riken} \affiliation{\rikjrbrc} 
\author{R.~Akimoto} \affiliation{\cns} 
\author{H.~Al-Bataineh} \affiliation{\nmsu} 
\author{J.~Alexander} \affiliation{\stonybrkc} 
\author{H.~Al-Ta'ani} \affiliation{\nmsu} 
\author{A.~Angerami} \affiliation{\columbia} 
% ...
\collaboration{PHENIX Collaboration} \noaffiliation

\date{\today}


\begin{abstract}

We report the measurement of cumulants ($C_n, n=1\ldots4$) of the 
net-charge distributions measured within pseudorapidity ($|\eta| <$ 0.35) 
in Au$+$Au collisions at $\sqrt{s_{_{NN}}}$=7.7--200 GeV with the PHENIX 
experiment at the Relativistic Heavy Ion Collider.  The ratios of 
cumulants (e.g. $C_1/C_2$, $C_3/C_1$) of the net-charge distributions, 
which can be related to volume independent susceptibility ratios, are 
studied as a function of centrality and energy. These quantities are 
important to understand the quantum-chromodynamics phase diagram and 
possible existence of a critical end point. The measured values are very 
well described by expectation from negative binomial distributions.  We do 
not observe any nonmonotonic behavior in the ratios of the cumulants as a 
function of collision energy.  The measured values of $C_1/C_2 = 
\mu/\sigma^2$ and $C_3/C_1 = S\sigma^3/\mu$ can be directly compared to 
lattice quantum-chromodynamics calculations and thus allow extraction of 
both the chemical freeze-out temperature and the baryon chemical potential 
at each center-of-mass energy.  The extracted baryon chemical potentials 
are in excellent agreement with a thermal-statistical analysis model.

\end{abstract}

	
\maketitle


One of the main goals in the study of relativistic heavy ion collisions is 
to map the quantum chromodynamics (QCD) phase diagram at finite 
temperature ($T$) and baryon chemical potential 
($\mu_B$)~\cite{Stephanov:1998dy}. Although the exact nature of the phase 
transition at finite baryon density is still not well established, several 
models suggest that, at large $\mu_B$ and low $T$, the phase transition 
between the hadronic phase and the quark-gluon-plasma (QGP) phase is of 
first order~\cite{Alford:1997zt,Stephanov:1996ki} and that at high $T$ and 
low $\mu_B$ there is a simple cross over from the QGP to hadronic 
phase~\cite{Aoki:2006we,Pisarski:1983ms,Stephanov:2004wx,Fodor:2004nz,Ejiri:2008xt}. 
The point at which the first-order phase transition ends in the $T-\mu_B$ 
plane is called the QCD critical end point (CEP), which is one of the 
central targets of the Relativistic Heavy Ion Collider (RHIC) 
beam-energy-scan program.  Several calculations also reported the possible 
existence of the CEP in the $T-\mu_B$ phase 
diagram~\cite{Fodor:2004nz,Stephanov:2004wx,Stephanov:1999zu}.

RHIC at Brookhaven National Laboratory has provided a large amount of data 
from \auau collisions at different colliding energies, which gives us a 
unique opportunity to scan the $T-\mu_B$ plane and investigate the 
possible existence and location of the CEP. In the thermodynamic limit, 
the correlation length ($\xi$) diverges at the 
CEP~\cite{Stephanov:1998dy}. Event-by-event fluctuations of various 
conserved quantities, such as net-baryon number, net-charge, and 
net-strangeness are proposed as possible signatures of the existence of 
the CEP~\cite{Koch:2005vg,Asakawa:2000wh,Asakawa:2009aj}. It has been 
shown in lattice QCD that with a next-to-leading-order Taylor series 
expansion around vanishing chemical potentials, the cumulants of 
charge-fluctuations are sensitive indicators for the occurrence of a 
transition from the hadronic to QGP phase~\cite{Ejiri:2005wq,Bazavov:2012vg}. 
Typically, the variances of net-baryon, net-charge, and net-strangeness 
distributions are proportional to $\xi$ as
$\sigma^2$(=$C_2$)=$\langle(\delta N)^2\rangle\sim\xi^2$~\cite{Stephanov:1999zu}, where $N$ is the 
multiplicity, ${\delta}N=N-\mu$ and $\mu$(=$C_1$) is the mean of the 
distribution.

Recent calculations reveal that higher cumulants of the fluctuations are 
much more sensitive to the proximity of the CEP than earlier 
measurements using second cumulants 
($\sigma^2$)~\cite{Stephanov:2008qz,Asakawa:2009aj}. The skewness ($S$) 
and kurtosis ($\kappa$) are related to the third and fourth moments $S$ (= 
$C_3/C_2^{3/2}) =\langle(\delta N)^3\rangle/\sigma^3$ $\sim \xi^{4.5}$ and 
$\kappa$(=$C_4/C_2^2)=\langle(\delta N)^4\rangle /\sigma^4 - 3 \sim 
\xi^7$. The ratio of the various order ($n$) of cumulants ($C_{n}$) and 
conventional values ($\mu$, $\sigma$, $S$ and $\kappa$)  can be related as 
follows: $\mu/\sigma^2$ = $C_{1}/C_{2}$, $S\sigma$ = $C_{3}/C_{2}$, 
$\kappa\sigma^2$ = $C_{4}/C_{2}$, and $S\sigma^3/\mu$ = $C_{3}/C_{1}$. 
Because $\xi$ diverges at the CEP, the ratios of cumulants $S\sigma$ and 
$\kappa\sigma^2$ should rise rapidly when approaching the 
CEP~\cite{Gavai:2010zn,Cheng:2008zh}. The cumulants of conserved 
quantities of net-baryon, net-charge, and net-strangeness obtained from 
lattice QCD calculations~\cite{Ejiri:2005wq,Bazavov:2012vg,Cheng:2008zh} 
and a hadron resonance gas (HRG) model~\cite{Karsch:2010ck} are related to 
the generalized susceptibilities of $n$-th order ($\chi^n$) associated 
with the conserved quantum numbers as $\mu/\sigma^2 \sim 
\chi^{(1)}/\chi^{(2)}$, $S\sigma \sim \chi^{(3)}/\chi^{(2)}$, 
$S\sigma^3/\mu \sim \chi^{(3)}/\chi^{(1)}$, and $\kappa\sigma^2 \sim 
\chi^{(4)}/\chi^{(2)}$. One advantage of measuring $\mu/\sigma^2$, 
$S\sigma$, $S\sigma^3/\mu$, and $\kappa\sigma^2$ is that the volume 
dependence of $\mu$, $\sigma$, $S$, and $\kappa$ cancel out in the ratios, 
hence theoretical calculations can be directly compared with the 
experimental measurements. These cumulant ratios can also be used to 
extract the freeze-out parameters and the location of the 
CEP~\cite{Bazavov:2012vg}. Net-electric charge fluctuations are more 
straightforward to measure experimentally than net-baryon number 
fluctuations, which are partially accessible via net-proton 
measurement~\cite{Aggarwal:2010wy}.  While net-charge fluctuations are not 
as sensitive as net-baryon fluctuations to the theoretical parameters, 
both measurements are desirable for a full understanding of the theory.

We report here precise measurements of the energy and centrality 
dependence of higher cumulants of net-charge multiplicity 
($\Delta N_{{\rm ch}}$ = $N^{+} - N^{-}$) distributions measured by the 
PHENIX experiment at RHIC in \auau collisions at \sqsn~=~7.7, 19.6, 27, 
39, 62.4, and 200 GeV.  These measurements cover a broad range of $\mu_B$ 
in the QCD phase diagram.

The PHENIX detector is composed of two central spectrometer arms, two 
forward muon arms, and global detectors~\cite{Adcox:2003zm}.  In this 
analysis, we use the central arm spectrometers, which cover a 
pseudorapidity range of $|\eta| \leq$ 0.35.  Each of the two arms subtends 
$\pi$/2 radians in azimuth and is designed to detect charged hadrons, 
electrons, and photons.  For data taken at \sqsn~=~62.4 and 200 GeV in 
2010 and 2007, respectively, the event centrality is determined using total 
charge deposited in the beam-beam counters (BBC), which are also used for 
triggering and vertex determination.  For lower energies (\sqsn~=~39 GeV 
and below) the acceptance of the BBCs (3.0 $< |\eta| <$ 3.9) are within 
the fragmentation region, so alternate detectors must be employed. For 
data taken at \sqsn~=~39 and 7.7 GeV in 2010, centrality is determined 
using the total charge deposited in the outer ring of the reaction plane 
detector (RXNP), which covers $1.0<|\eta|<1.5$~\cite{Richardson:2010hm}. 
For data taken at \sqsn~=~19.6 and 27 GeV in 2011, the RXNP was absent, 
so centrality is determined using the total energy of electromagnetic 
calorimeter (EMCal) clusters to minimize the correlation with the 
charge of the tracks measured in the same acceptance.  More details on the 
procedure are given in~\cite{Adler:2004zn}. The analyzed events for the 
above mentioned energies are within a collision vertex of 
$|Z_{\rm vertex}|<$ 30 cm. The number of analyzed events are 2M, 6M, 21M, 
154M, 474M, and 1681M for \sqsn = 7.7, 19.6, 27, 39, 62.4, and 200 GeV 
\auau collisions, respectively.

The number of positively charged ($N^{+}$) and negatively charged 
($N^{-}$) particles measured on an event-by-event basis are used to 
calculate the net-charge ($\Delta N_{{\rm ch}}$) distributions for each 
collision centrality and energy. The charged-particle trajectories are 
reconstructed using information from the drift chamber and pad chambers 
(PC1 and PC3). A combination of reconstructed drift-chamber tracks and 
matching hits in PC1 are used to determine the momentum and charge of the 
particle. Tracks having a transverse momentum (\pt) between 0.3 and 2.0 
GeV/$c$ are selected for this analysis.  The ring imaging \v{C}erenkov 
detector is used to reduce the electron background resulting from 
conversion photons. To further reduce the background, selected tracks are 
required to lie within a 2.5$\sigma$ matching window between track 
projections and PC3 hits, and a 3$\sigma$ matching window for the EMCal.

\begin{figure}[ht]
\centering
\includegraphics[width=1.0\linewidth]{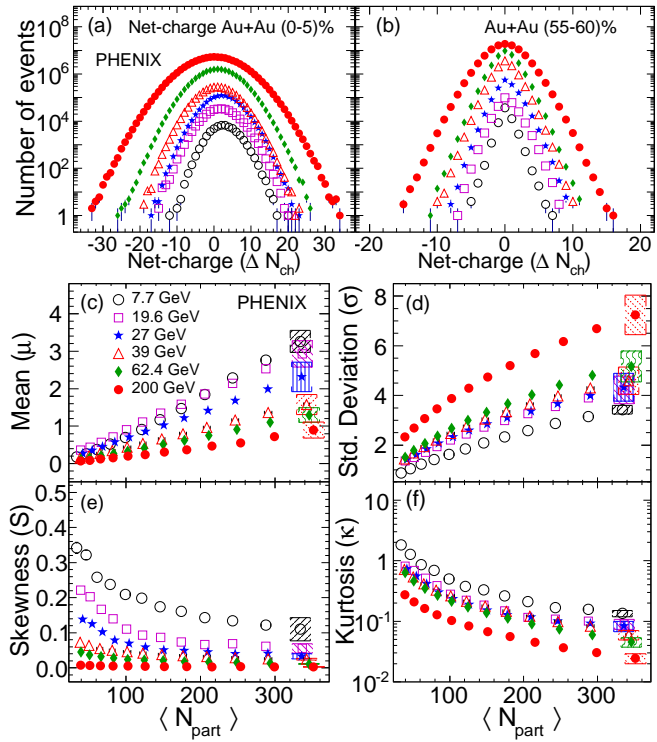}
\caption{(Color online). 
Uncorrected net-charge ($\Delta N_{{\rm ch}}$) distributions, within 
$|\eta| \le$ 0.35 for different energies, from \auau collisions for (a) 
central (0\%--5\%) and (b) peripheral (55\%--60\%) centrality.  (c)--(f) 
are the efficiency corrected cumulants of net-charge distributions as a 
function of $\langle \Npart \rangle$ from \auau collisions at different 
collision energies.  Systematic uncertainties on moments are shown for 
central (0\%--5\%) collisions.}
\label{fig:fig1}
\end{figure}

Figure~\ref{fig:fig1}(a) and (b) show $\Delta$\Nch distributions 
in \auau collisions for central (0\%--5\%) and peripheral (55\%--60\%) 
collisions at different collision energies.  These $\Delta$\Nch 
distributions are not corrected for reconstruction efficiency. The 
centrality classes associated with the average number of participants 
($\langle \Npart \rangle$) are defined for each 5\% centrality bin. These 
classes are determined using a Monte-Carlo simulation based on Glauber 
model calculations with the BBC, RXNP, and EMCal detector response taken 
into account~\cite{Adler:2004zn,Miller:2007ri}.

The $\Delta\Nch$ distributions are characterized by cumulants and related 
quantities such as $\mu$, $\sigma$, $S$, and $\kappa$, which are 
calculated 
from the distributions. The statistical uncertainties for the cumulants 
are calculated using the bootstrap method~\cite{Bootstrap}. Corrections 
are then made for the reconstruction efficiency, which is estimated for 
each centrality and energy using the {\sc hijing}1.37 event 
generator~\cite{Wang:1991hta} and then processed through a {\sc geant} 
simulation with the PHENIX detector setup. For all collision energies, the 
average efficiency for detecting the particles within the acceptance 
varies between 65\%--72\% and 76\%--85\% for central (0\%--5\%) and 
peripheral (55\%--60\%) events, respectively with 4\%--5\% variation as a 
function of energy. The efficiency correction applied to the cumulants is 
based on a binomial probability distribution for the reconstruction 
efficiency~\cite{Bzdak:2012an}. The efficiency corrected $\mu$, $\sigma$, 
$S$, and $\kappa$ as a function of $\langle \Npart \rangle$ are shown in 
panels (c-f) of Fig.~\ref{fig:fig1}.

The $\mu$ and $\sigma$ for net-charge distributions increase with 
increasing $\langle\Npart\rangle$, while $S$ and $\kappa$ decrease with 
increasing $\langle\Npart\rangle$ for all collision energies. At a given 
$\langle\Npart\rangle$ value, $\mu$, $S$, and $\kappa$ of net-charge 
distributions decrease with increasing collision energy. However, the 
width ($\sigma$) of net-charge distributions increases with increasing 
collision energy indicating the increase of fluctuations in the system at 
higher \sqsn.

\begin{figure}[ht]
\centering
\includegraphics[width=1.0\linewidth]{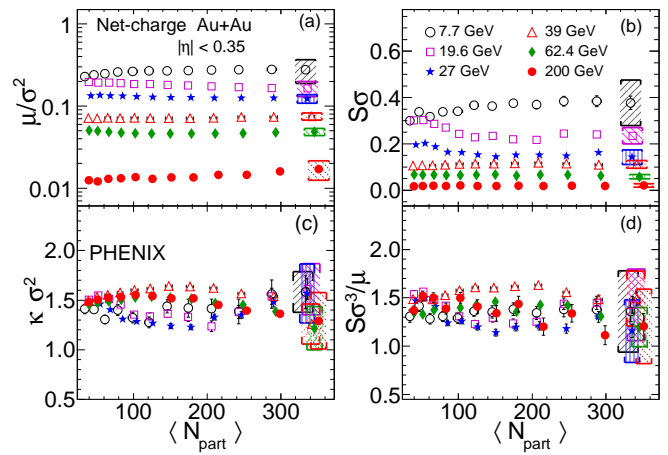}
\caption{(Color online). 
$\langle \Npart \rangle$ dependence of efficiency corrected (a) 
$\mu/\sigma^2$, (b) $S\sigma$, (c) $\kappa\sigma^2$, and (d) 
$S\sigma^3/\mu$ of net-charge distributions for \auau collisions at 
different collision energies. Statistical errors are shown along with the 
data points while systematic uncertainties are shown for (0\%--5\%) 
collisions.}
\label{fig:fig2}
\end{figure}

The systematic uncertainties are estimated by: (1) varying the $Z_{{\rm 
vertex}}$ cut to less than $\pm$10 cm; (2) varying the matching parameters 
of PC3 hits and EMCal clusters with the projected tracks to study the 
effect of background tracks originating from secondary interactions or 
from ghost tracks; (3) varying the centrality bin width to study 
nondynamical contributions to the net-charge fluctuations due to the 
finite width of the centrality 
bins~\cite{Adare:2008ns,Adler:2007fj,Konchakovski:2006aq}; and (4) varying 
the lower \pt cut. The total systematic uncertainties estimated for 
various cumulants for all energies are: 10\%--24\% for $\mu$, 5\%--10\% 
for $\sigma$, 25\%--30\% for $S$, and 12\%--19\% for $\kappa$. The 
systematic uncertainties are similar for all centralities at a given 
energy and are treated as uncorrelated as a function of \sqsn. For clarity 
of presentation, the systematic uncertainties are only shown for central 
(0\%--5\%) collisions.

\begin{figure}[ht]
\centering
\includegraphics[width=1.0\linewidth]{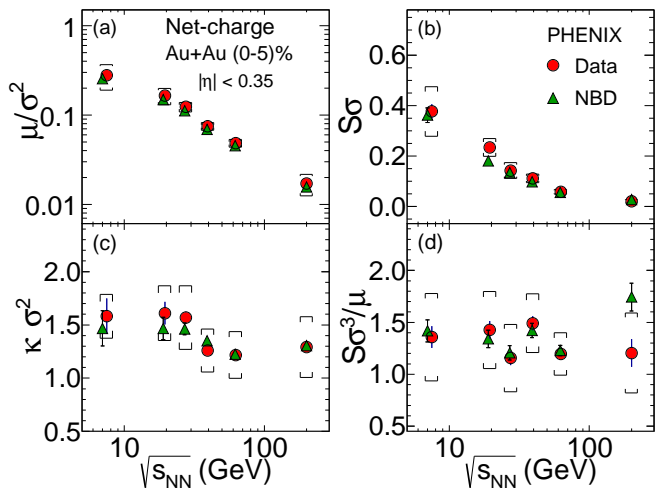}
\caption{(Color online). 
The energy dependence of efficiency corrected (a) $\mu/\sigma^2$, (b) 
$S\sigma$, (c) $\kappa\sigma^2$, and (d) $S\sigma^3/\mu$ of net-charge 
distributions for central (0\%--5\%) \auau collisions. The error bars are 
statistical and caps are systematic uncertainties. The triangle symbol 
shows the corresponding efficiency corrected cumulant ratios for 
net-charge, from NBD fits to the individual $N^+$ and $N^-$ 
distributions.}
\label{fig:fig3}
\end{figure}

Figure~\ref{fig:fig2} shows the $\langle \Npart \rangle$ dependence of 
$\mu/\sigma^2$, $S\sigma$, $\kappa\sigma^2$, and $S\sigma^3/\mu 
(=(S\sigma)/(\mu/\sigma^2))$ extracted from the net-charge distributions 
in \auau collisions at different \sqsn. The results are corrected for the 
reconstruction efficiencies. Statistical uncertainties are shown along 
with the data points. The systematic uncertainties are constant fractional 
errors for all centralities at a particular energy, hence they are 
presented for the central (0\%--5\%) collision data point only. The 
systematic uncertainties on these ratios across different energies varies 
as follows: 20\%--30\% for $\mu/\sigma^2$, 15\%--34\% for $S\sigma$, 
12\%--22\% for $\kappa\sigma^2$, and 17\%--32\% for $S\sigma^3/\mu$. It is 
observed in Fig.~\ref{fig:fig2} that the ratios of the cumulants are 
weakly dependent on $\langle \Npart \rangle$ for each collision energy; 
the values of $\mu/\sigma^2$ and $S\sigma$ decrease from lower to higher 
collision energies, while the $\kappa\sigma^2$ and $S\sigma^3/\mu$ values 
are constant as a function of \sqsn within systematic uncertainties.

The collision energy dependence of $\mu/\sigma^2$, $S\sigma$, 
$\kappa\sigma^2$ and $S\sigma^3/\mu$ of the net-charge distributions for 
central (0\%--5\%) \auau collisions are shown in Fig.~\ref{fig:fig3}. The 
statistical and systematic uncertainties are shown along with the data 
points. The experimental data are compared with 
negative-binomial-distribution (NBD) expectations, which are calculated by 
computing the efficiency corrected cumulants for the measured $N^+$ and 
$N^-$ distributions fit with NBD's respectively, which also describe total 
charge ($N^+ + N^-$) distributions very 
well~\cite{Adare:2008ns,Adler:2007fj}. The various order ($n$ = 1, 2, 3 
and 4) of net-charge cumulants from NBD are given as $C_n(\Delta N_{{\rm 
ch}})$ = $C_n(N^+) + (-1)^n C_n(N^-)$, where $C_n(N^+)$ and $C_n(N^-)$ are 
cumulants of $N^+$ and $N^-$ distributions, 
respectively~\cite{Neilsen,Tarnowsky:2012vu}.

The $\mu/\sigma^2$ and $S\sigma$ values in Fig.~\ref{fig:fig3}(a) and 
Fig.~\ref{fig:fig3}(b), respectively both decrease with increasing \sqsn. 
The NBD expectation agrees well with the data. The $\kappa\sigma^2$ values 
in Fig.~\ref{fig:fig3}(c) remain constant and positive, between $1.0 < 
\kappa\sigma^2 < 2.0$ at all the collision energies within the statistical 
and systematic uncertainties. However, there is $\sim$ 25\% increase of 
$\kappa\sigma^2$ values at lower energies compared to higher energies 
above \sqsn = 39 GeV, which is within the systematic uncertainties. These 
data are in agreement with a previous measurement~\cite{Adamczyk:2014fia}, 
but provide a more precise determination of the higher cumulant ratios, 
verified by the NBD method of correcting for efficiency, which is simple 
and analytical for all cumulant ratios with the standard binomial 
correction~\cite{Bzdak:2012an}. The $S\sigma^3/\mu$ values in 
Fig.~\ref{fig:fig3}(d) remain constant at all collision energies within 
the uncertainties and are well described by the NBD expectation. From the 
energy dependence of $\mu/\sigma^2$, $S\sigma$, $\kappa\sigma^2$, and 
$S\sigma^3/\mu$, no obvious nonmonotonic behavior is observed.
Although both previous measurements by 
STAR~\cite{Adamczyk:2014fia,Adamczyk:2013dal} use the pseudorapidity range 
$|\eta|\leq0.5$, compared to the present measurement spanning $|\eta|\leq 
0.35$, these measurements are all within the central rapidity region and 
are expected to be valid for comparison to lattice QCD calculations. The 
efficiency corrected results for the cumulant ratios $\mu/\sigma^2$, 
$S\sigma$, and $\kappa\sigma^2$ remain the same within statistics whether 
each single arm of the PHENIX central spectrometer (azimuthal aperture 
$\delta\phi = \pi/2$) or both arms ($\delta\phi = \pi$) are used. This is 
a clear verification of the insensitivity of measured cumulant ratios to 
volume effects.

\begin{figure}[hbt]
\includegraphics[width=1.0\linewidth]{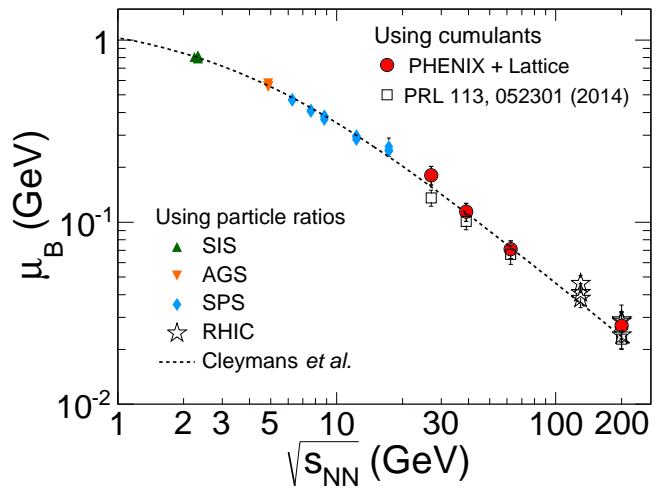}
\caption{(Color online). 
The energy dependence of the chemical freeze-out parameter $\mu_B$. The 
dashed line is the parametrization given in Ref.~\cite{Cleymans:2005xv} 
and the other experimental data are from Ref.~\cite{Cleymans:2005xv} and 
references therein.
}
\label{fig:fig4}
\end{figure}

\begin{table*}[hbt]
 \caption{
\label{tbl:table1}
Freeze-out $T_f$ and $\mu_B$ vs. $\sqsn$ in the range $27\leq \sqsn\leq 
200$ GeV from this work compared to $\mu_B$ values from 
Ref.~\cite{Borsanyi:2014ewa}, which used STAR net-charge cumulant 
measurements from Ref.~\cite{Adamczyk:2014fia} for $\mu_B$; with $140$ MeV 
$\leq T_f\leq150$ MeV obtained from STAR net-proton measurement in 
Ref.~\cite{Adamczyk:2013dal} by averaging $S\sigma^3/\mu$ over \sqsn = 27, 
39, 62.4 and 200 GeV.}
\begin{ruledtabular} \begin {tabular}{cccccc}
      & \multicolumn{2}{c}{PHENIX 
+ Ref.~\cite{Bazavov:2012vg, Mukherjee}}
& \multicolumn{2}{c}{PHENIX + 
Ref.~\cite{Borsanyi:2013hza}} 
& STAR + Ref.~\cite{Borsanyi:2014ewa}\\
\sqsn (GeV) 
& $T_f$ (MeV)& $\mu_B$ (MeV) 
& $T_f$ (MeV)& $\mu_B$ (MeV) 
& $\mu_B$ (MeV)\\
\hline
27
&   $164\pm6$ 
&$181\pm21$ 
&$160\pm6$  
&$184\pm21$ 
& $136\pm13.8$\\
39
&   $158\pm5$ 
&$114\pm13$ 
&$156\pm5$  
&$118\pm10$ 
& $101\pm10$\\
62.4
& $163\pm5$ 
&$71\pm8$   
&$159\pm5$  
&$74\pm8$   
& $66.6\pm7.9$\\
200
&  $163\pm8$ 
&$27\pm5$   
&$159\pm8$  
&$25\pm7$   
& $22.8\pm2.6$\\
\end{tabular}   \end{ruledtabular} 
\end{table*}

The precise measurement of both $\mu/\sigma^2$ and $S\sigma^3/\mu$ in the 
present study allow both $\mu_B$ and $T_f$ to be determined, unlike a 
previous calculation in Ref.~\cite{Borsanyi:2014ewa,Borsanyi:2013hza}, which 
was only able to use the $\mu/\sigma^2$ measurement from 
Ref.~\cite{Adamczyk:2014fia}. The comparison of $S\sigma^{3}/\mu$ for different 
\sqsn with the lattice calculations (Fig. 3(b) in 
Ref.~\cite{Bazavov:2012vg,Mukherjee}) enables us to extract the chemical 
freeze-out temperature ($T_{f}$). Furthermore, $\mu_B$ can be extracted by 
comparing the measured $\mu/\sigma^2$ ratios with the lattice calculations of 
$R_{12} = \mu/\sigma^2$ (Fig.3(a) in Ref.~\cite{Bazavov:2012vg,Mukherjee}). The 
extracted $T_{f}$ and $\mu_B$ values are listed in Table~\ref{tbl:table1}. 
The $T_{f}$ and $\mu_B$ extracted using the lattice calculations in 
the continuum limit from Ref.~\cite{Borsanyi:2013hza} are also depicted 
in Table~\ref{tbl:table1}.  The extracted freeze-out parameters using 
different lattice results agree very well. However, the extracted $T_f$ 
are 2-4 MeV lower using Ref.~\cite{Borsanyi:2013hza} than with 
Ref.~\cite{Bazavov:2012vg,Mukherjee}, which is well within the stated 
uncertainties.
The detailed freeze-out parameter extraction procedure is given 
in Ref.~\cite{Bazavov:2012vg,Borsanyi:2013hza,Borsanyi:2014ewa}. This is a 
direct combination of experimental data and lattice calculations to extract 
physical quantities. The \sqsn dependence of $\mu_B$ shown in 
Fig.~\ref{fig:fig4} is in agreement with the thermal-statistical analysis 
model of identified particle yields~\cite{Cleymans:2005xv}. The $\mu_B$ extracted 
in the present net-charge measurement and the values reported 
in~\cite{Borsanyi:2014ewa} are in agreement within stated uncertainties, with 
some tension at \sqsn = 27 GeV. Available lattice results allow extraction of 
$\mu_B$ and $T_f$ from \sqsn~=~27 GeV and higher using the present net-charge 
experimental data. Other recent 
calculations~\cite{Alba:2014eba,Bazavov:2015zja} have used both net-proton 
and net-charge measurements to estimate the freeze-out parameters.

In summary, fluctuations of net-charge distributions have been studied 
using higher cumulants ($\mu$, $\sigma$, $S$, and $\kappa$) for $|\eta| <$ 
0.35 with the PHENIX experiment in \auau collisions ranging from \sqsn = 
7.7 to 200 GeV. The ratios of cumulants ($\mu/\sigma^2$, $S\sigma$, 
$\kappa\sigma^2$, and $S\sigma^3/\mu$) have been derived from the 
individual cumulants of the distributions studied as a function of 
$\langle\Npart\rangle$ and \sqsn. The $\mu/\sigma^2$ and $S\sigma$ values 
decrease with increasing collision energy and are weakly dependent on 
centrality, whereas $\kappa\sigma^2$ and $S\sigma^3/\mu$ values remain 
constant over all collision energies within uncertainties. The efficiency 
corrected values from the NBD expectation reproduce the experimental data. 
These data are in agreement with a previous 
measurement~\cite{Adamczyk:2014fia}, but provide more precise determination of 
the higher cumulant ratios $S\sigma$ and $\kappa\sigma^2$. In the present study 
we do not observe any significant nonmonotonic behavior of $\mu/\sigma^2$, 
$S\sigma$, $\kappa\sigma^2$, and $S\sigma^3/\mu$ as a function of collision 
energies. Comparison of the present measurements together with the lattice 
calculations enables us to extract the freeze-out temperature $T_f$ and baryon 
chemical potential ($\mu_B$) over a range of collision energies. The extracted 
$\mu_B$ values are in excellent agreement with the thermal-statistical analysis 
model~\cite{Cleymans:2005xv}.




We thank the staff of the Collider-Accelerator and Physics
Departments at Brookhaven National Laboratory and the staff of
the other PHENIX participating institutions for their vital
contributions.  We thank F. Karsch and S. Mukherjee for providing us with 
tables of their calculations and for helpful discussions. We acknowledge 
support from the Office of Nuclear Physics in the Office of Science of the 
Department of Energy, the National Science Foundation, Abilene Christian 
University Research Council, Research Foundation of SUNY, and Dean of the 
College of Arts and Sciences, Vanderbilt University (U.S.A),
Ministry of Education, Culture, Sports, Science, and Technology
and the Japan Society for the Promotion of Science (Japan),
Conselho Nacional de Desenvolvimento Cient\'{\i}fico e
Tecnol{\'o}gico and Funda\c c{\~a}o de Amparo {\`a} Pesquisa do
Estado de S{\~a}o Paulo (Brazil),
Natural Science Foundation of China (People's Republic of China),
Ministry of Science, Education, and Sports (Croatia),
Ministry of Education, Youth and Sports (Czech Republic),
Centre National de la Recherche Scientifique, Commissariat
{\`a} l'{\'E}nergie Atomique, and Institut National de Physique
Nucl{\'e}aire et de Physique des Particules (France),
Bundesministerium f\"ur Bildung und Forschung, Deutscher
Akademischer Austausch Dienst, and Alexander von Humboldt Stiftung 
(Germany),
National Science Fund, OTKA, K\'aroly R\'obert University College,
and the Ch. Simonyi Fund (Hungary),
Department of Atomic Energy and Department of Science and Technology 
(India),
Israel Science Foundation (Israel),
Basic Science Research Program through NRF of the Ministry of Education 
(Korea),
Physics Department, Lahore University of Management Sciences (Pakistan),
Ministry of Education and Science, Russian Academy of Sciences,
Federal Agency of Atomic Energy (Russia),
VR and Wallenberg Foundation (Sweden),
the U.S. Civilian Research and Development Foundation for the
Independent States of the Former Soviet Union,
the Hungarian American Enterprise Scholarship Fund,
and the US-Israel Binational Science Foundation.



%
 
\end{document}